\documentclass[%
 reprint,
superscriptaddress,
 amsmath,amssymb,
 aps,
]{revtex4-2}
\usepackage{graphicx}% Include figure files
\usepackage{dcolumn}% Align table columns on decimal point
\usepackage{bm}% bold math
\usepackage{comment}
\usepackage{hyperref}% add hypertext capabilities
\usepackage{braket}

\usepackage{hyperref}
\usepackage{xcolor}
\definecolor{cite}{rgb}{0.,0.,0.9}  
\hypersetup{colorlinks,linkcolor={cite},citecolor={cite},urlcolor={cite}}

\begin{document}

%\preprint{APS/123-QED}

\title{A Portable Dual-Color Two-Photon Rubidium Optical Frequency Standard}

\author{Sarah K. Scholten}
\email{sarah.scholten@adelaide.edu.au}
\affiliation{
 Institute for Photonics, Advanced Sensing and Quantum Technologies (IPAS-QT) and School of Physics, Chemistry and Earth Sciences, Adelaide University, Adelaide SA 5005, Australia
}
\affiliation{
 ARC Centre of Excellence in Optical Microcombs for Breakthrough Science (COMBS)
}
\affiliation{
Collaborative Research Hub for Alternative Positioning,
Navigation and Timing (aPNT), Adelaide University, Adelaide SA 5005 
}

\author{Emily Ahern}
\affiliation{
 Institute for Photonics, Advanced Sensing and Quantum Technologies (IPAS-QT) and School of Physics, Chemistry and Earth Sciences, Adelaide University, Adelaide SA 5005, Australia
}

\author{Clayton Locke}
\affiliation{
 Institute for Photonics, Advanced Sensing and Quantum Technologies (IPAS-QT) and School of Physics, Chemistry and Earth Sciences, Adelaide University, Adelaide SA 5005, Australia
}
%\affiliation{
%QuantX Labs, Level 2, SpaceLab Building Frome Road, Lot Fourteen, Adelaide SA 5000
%}

\author{Christopher J. Billington}
\affiliation{
 Institute for Photonics, Advanced Sensing and Quantum Technologies (IPAS-QT) and School of Physics, Chemistry and Earth Sciences, Adelaide University, Adelaide SA 5005, Australia
}
\affiliation{
Collaborative Research Hub for Alternative Positioning,
Navigation and Timing (aPNT), Adelaide University, Adelaide SA 5005 
}

\author{Nicolas Bourbeau H\'{e}bert}
\affiliation{
 Institute for Photonics, Advanced Sensing and Quantum Technologies (IPAS-QT) and School of Physics, Chemistry and Earth Sciences, Adelaide University, Adelaide SA 5005, Australia
}
\affiliation{
Collaborative Research Hub for Alternative Positioning,
Navigation and Timing (aPNT), Adelaide University, Adelaide SA 5005 
}

\author{Montana Nelligan}
\affiliation{
 Institute for Photonics, Advanced Sensing and Quantum Technologies (IPAS-QT) and School of Physics, Chemistry and Earth Sciences, Adelaide University, Adelaide SA 5005, Australia
}

\author{Ashby P. Hilton}
\affiliation{
 Institute for Photonics, Advanced Sensing and Quantum Technologies (IPAS-QT) and School of Physics, Chemistry and Earth Sciences, Adelaide University, Adelaide SA 5005, Australia
}
\affiliation{
Collaborative Research Hub for Alternative Positioning,
Navigation and Timing (aPNT), Adelaide University, Adelaide SA 5005 
}

\author{Rachel F. Offer}
\affiliation{
 Institute for Photonics, Advanced Sensing and Quantum Technologies (IPAS-QT) and School of Physics, Chemistry and Earth Sciences, Adelaide University, Adelaide SA 5005, Australia
}

\author{Elizaveta Klantsataya}
\affiliation{
 Institute for Photonics, Advanced Sensing and Quantum Technologies (IPAS-QT) and School of Physics, Chemistry and Earth Sciences, Adelaide University, Adelaide SA 5005, Australia
}
\affiliation{
Collaborative Research Hub for Alternative Positioning,
Navigation and Timing (aPNT), Adelaide University, Adelaide SA 5005 
}

\author{Christopher Perrella}
\affiliation{
 Institute for Photonics, Advanced Sensing and Quantum Technologies (IPAS-QT) and School of Physics, Chemistry and Earth Sciences, Adelaide University, Adelaide SA 5005, Australia
}
\affiliation{
 ARC Centre of Excellence in Optical Microcombs for Breakthrough Science (COMBS)
}
\affiliation{
 Centre of Light for Life and School of Biological Sciences, Adelaide University, Adelaide, South Australia, 5005, Australia
}

\author{Andre N. Luiten}
\affiliation{
 Institute for Photonics, Advanced Sensing and Quantum Technologies (IPAS-QT) and School of Physics, Chemistry and Earth Sciences, Adelaide University, Adelaide SA 5005, Australia
}
\affiliation{
 ARC Centre of Excellence in Optical Microcombs for Breakthrough Science (COMBS)
}
%\affiliation{
%QuantX Labs, Level 2, SpaceLab Building Frome Road, Lot Fourteen, Adelaide SA 5000
%}
\affiliation{
Collaborative Research Hub for Alternative Positioning,
Navigation and Timing (aPNT), Adelaide University, Adelaide SA 5005 
}

\date{\today}

%%%%%%%%%%%%%%%%%%%%%%%%%%%%%%%%%%%%%%%%%%%%%%%%%%%%%%%%%%%%%%%%%%%%%%%%%%%%%%%%%%%%%%%%%%%%%%%%%%%
%-------------------------------------------------------------------------------------------------%
%%%%%%%%%%%%%%%%%%%%%%%%%%%%%%%%%%%%%%%%%%%%%%%%%%%%%%%%%%%%%%%%%%%%%%%%%%%%%%%%%%%%%%%%%%%%%%%%%%%

\begin{abstract}
Portable atomic clocks are essential in a wide variety of applications, most notably in the operation of global navigation satellite systems.
%Existing portable atomic clocks utilizing microwave-based interrogation schemes are now routinely eclipsed by the next generation of atomic frequency standards based on optical interrogation.
%While optical frequency standards demonstrate greatly improved frequency stability, they have only recently reached a level of technical maturity required demonstrate this improved performance outside of well curated laboratory environments.
Existing portable atomic clocks utilizing microwave-based interrogation schemes are now routinely eclipsed by the next generation of atomic frequency standards based on optical interrogation.
While optical frequency standards demonstrate greatly improved frequency stability, they have only recently reached a level of technical maturity required to demonstrate this improved performance outside of well curated laboratory environments.
Here, we demonstrate a fully autonomous and portable optical frequency standard based on an efficient dual-color excitation of the $5S_{1/2}\rightarrow5D_{5/2}$ two-photon transition in $^{87}$Rb.
The standard utilizes a combination of robust, highly developed commercial-off-the-shelf telecommunications technologies and a fully integrated portable optical frequency comb, providing the optical and microwave outputs vital for interfacing with existing electronic systems and infrastructure.
The system demonstrates a fractional frequency stability of $1.9\times10^{-13}$ at 1\,s of integration time, reaching $3.5\times10^{-15}$ at 8000\,s of integration time without the need for drift removal.
This portable demonstrator unit marks a significant achievement in the development of Rb optical atomic frequency standards, and for the deployment of optical atomic frequency standards outside of the laboratory. 
\end{abstract}

\maketitle

%\tableofcontents

%%%%%%%%%%%%%%%%%%%%%%%%%%%%%%%%%%%%%%%%%%%%%%%%%%%%%%%%%%%%%%%%%%%%%%%%%%%%%%%%%%%%%%%%%%%%%%%%%%%
%-------------------------------------------------------------------------------------------------%
%%%%%%%%%%%%%%%%%%%%%%%%%%%%%%%%%%%%%%%%%%%%%%%%%%%%%%%%%%%%%%%%%%%%%%%%%%%%%%%%%%%%%%%%%%%%%%%%%%%

\section{Introduction} \label{sec:Intro}

Atomic clocks are vital to position, navigation, and timing services for both civilian and military use globally~\cite{Bandi2023}. 
They provide the timing basis for global navigation satellite systems (GNSS) such as the global positioning system (GPS)~\cite{Batori2020,Schuldt2020,Hollberg2021, Jaduszliwer2021}. 
The past decade has seen increased interest in atomic clocks for use in Earth-based GPS-denied environments, with a worldwide effort to develop portable atomic clocks with stabilities surpassing those on the GNSS~\cite{Burt2021, Roslund2024, Martin2018,Newman2021, Lemke}. 

A portable atomic clock based on a microwave transition in mercury was demonstrated in low Earth orbit for two months~\cite{Burt2021}, achieving a fractional frequency stability of $2{\times}10^{-13}$ at an integration time ($\tau$) of $100$\,s, and $3{\times}10^{-15}$ at $\tau\,{=}\,23$\,days. 
In 2019, an iodine-based optical frequency standard was demonstrated on a sounding rocket~\cite{Doringshoff2019}.
In the six minute flight time, a fractional frequency stability of $2{\times}10^{-13}$ at $\tau\,{=}\,100$\,s was achieved. 
In 2022, Vector Atomic demonstrated an iodine-based optical frequency standard at sea in the Rim of the Pacific (RIMPAC) international maritime exercise~\cite{Roslund2024}. 
This frequency standard demonstrated  $5{\times}10^{-14}$ at $\tau\,{=}\,1$\,s and  $4{\times}10^{-15}$ at $\tau\,{=}\,100$\,s performance.

\begin{figure*}[t]
\includegraphics[width=\textwidth]{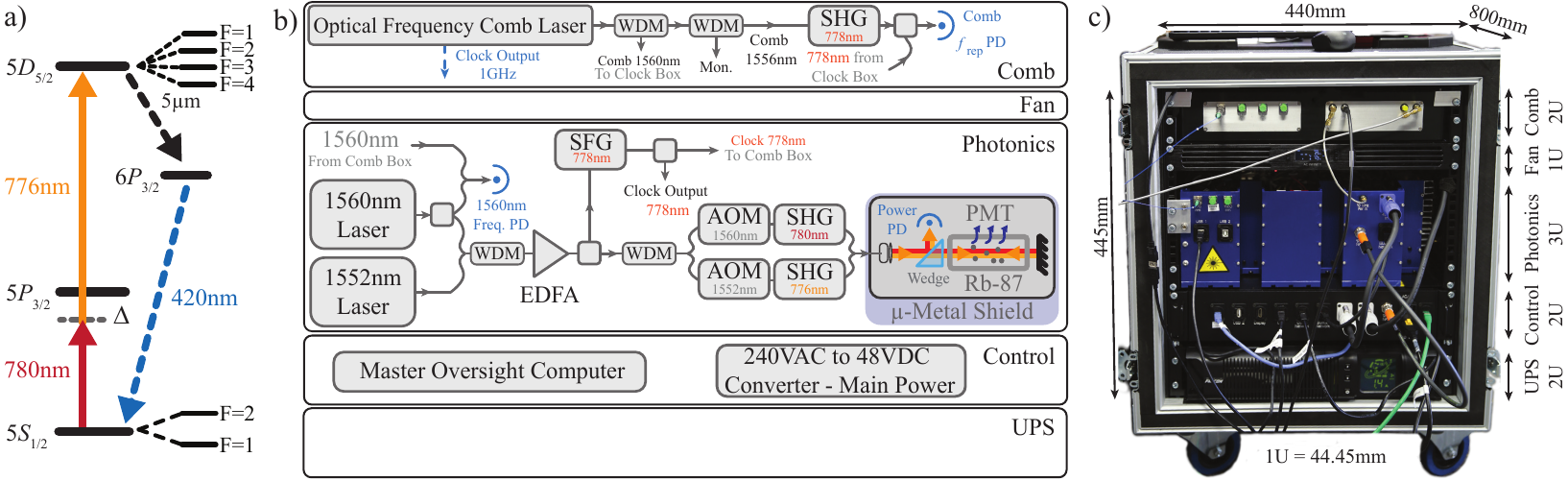}
\caption{\label{fig:ClockInManyBoxes} 
a) Dual-color excitation of the $5S_{1/2}\rightarrow5D_{5/2}$ clock transition in the two-photon frequency standard: one 776\,nm and 780\,nm photon together excite the transition; successful excitation is evidenced by production of 420\,nm fluorescence from the 6$P_{3/2}$ state decay pathway. 
b) Portable frequency standard subsystems are housed in interconnected housings with substantial air spacing for thermal management in challenging field conditions. 
%From top to bottom: 2U portable optical frequency comb; 1U commercial fan unit; the 3U frequency standard photonics, smaller power supplies and local control systems (`Photonics'); 2U Control box housing the master oversight computer; the 2U UPS unit.
In the Comb and Photonics housings, optical components are shown in grey with solid lines; electrical signals are shown in blue dashed lines.
WDM: wavelength division multiplexer; EDFA: erbium-doped fiber amplifier, AOM: acousto-optic modulator, SHG: second harmonic generator, SFG: sum-frequency generator, PMT: photo-multiplier tube, PD: photodetector, RF: radio-frequency.
c) Photo of the 10U 19-inch rack-mounted Rb frequency standard within the portable housing, dimensions shown.}
\end{figure*}

Another promising atomic candidate for potential realization of a compact frequency standard is $^{87}$Rb, with narrow multi-photon clock transitions easily accessible by robust frequency-doubled fiber lasers developed for the telecommunications industry.
Here, we describe the fully packaged, portable, autonomous Rb optical atomic frequency standard~\cite{ahern2024} that was field tested at the 2022 Rim of the Pacific exercise (RIMPAC-2022), which saw three portable clock architectures fielded for the first time~\cite{Roslund2024,ahern2024}.
Our frequency standard is based on the $5S_{1/2}\,(F=2)\rightarrow5D_{5/2}\,(F=4)$ two-photon transition of $^{87}$Rb.
This transition can be excited in a thermal vapor via a single-color scheme (two 778.1\,nm photons)~\cite{Martin2018, Newman2021, Lemke, Andeweg2026, Feng2026} or, as in this frequency standard, a dual-color scheme using one 780\,nm photon and one 776\,nm photon~\cite{ahern2024, EmilyPaper, Perrella2019}. 
Successful excitation in either scheme is evidenced by 420\,nm fluorescence that is produced via the $6P_{3/2}$ decay path as seen in Fig.\,\ref{fig:ClockInManyBoxes}a.
The single-color excitation scheme sees a large frequency detuning from the $5P_{3/2}$ intermediate state ($\Delta_i=1$\,THz).
The dual-color approach, in contrast, allows for the full control and tunability of $\Delta_i$~\cite{EmilyPaper}.
This allows the dual-color frequency standard to utilize tenfold less interrogating laser intensities while maintaining a similar transition rate ($W$) to the single-color approach:
\begin{equation}\label{eqn:TransRate}
    W \propto \frac{N}{\Gamma_e}\frac{I_1 I_2}{4 \Delta_i^2 + \Gamma_i}
\end{equation}
where $N$ is the atomic number density, $\Gamma_i$ and $\Gamma_e$ are the decay rates of the ground to intermediate state transition, and intermediate to excited state transition respectively, and $I_1$ and $I_2$ are the intensities of the two lasers driving the transition~\cite{Bjorkholm1976, ahern2024, EmilyPaper}. 
In the case of the $5S_{1/2}\,{\rightarrow}\,5D_{5/2}$ two-photon transition, the intermediate state is the $5P_{3/2}$ state and the excited state is the $5D_{5/2}$ state.
This work has a nominal $\Delta_i$ in the range of a few GHz, however the utilization of two colors introduces residual Doppler broadening ($\Gamma_e$) of approximately 4\,MHz~\cite{Bjorkholm1976, Perrella2019}. 
Nonetheless, the benefits of the increased transition rate and tenfold lowered laser intensity requirements with reduced $\Delta_i$ outweigh this broadening~\cite{EmilyPaper}.
%The additional degree of freedom introduced by a variable intermediate state detuning allows for a choice of compromise between higher transition rates and lower intensities, balanced against a larger residual Doppler broadening, increased light shifts, and spontaneous absorption by the intermediary state. 
%With this in mind, we have chosen to operate at a $\Delta_i$ of a few GHz, guided by the work in Ref.~\cite{EmilyPaper}. This results in a linewidth of ~700kHz, and an increase in scattering rate over the single-colour approach of BLAH for the same intensities. This allows us to instead reduce the interrogation intensities dramatically and still porduce a modest flux of fluoresence.

%In the ideal case, the short-term stability limit for such a frequency standard is set by the transition frequency ($\omega$), the spectral linewidth ($\Gamma_e$), and the signal-to-noise ratio (SNR) of the measurement of the transition, which is proportional in itself to $W$. This results in a fractional frequency stability:
%\begin{equation}
%    \sigma(\tau) \propto \frac{\Gamma_e}{\omega} \frac{1}{\text{SNR}} \frac{1}{\sqrt{\tau}}.
%\end{equation}
%where $\tau$ is the measurement integration time~\cite{Margolis2010}.

Previously published results of this dual-color scheme have demonstrated stability results of $1.6\,{\times}\,10^{-13}$ at $\tau=1$\,s and $8.6\,{\times}\,10^{-15}$ at $\tau=2000$\,s~\cite{EmilyPaper}.
In this work, we detail the technical design and demonstrate the performance of a fully packaged, portable, and automated version of this frequency standard architecture against other well-characterized laboratory standards.
This unit that is capable of unsupervised operation outside of the laboratory environment and was fielded at RIMPAC-2022 - the same exercise as Ref ~\cite{Roslund2024}, with extra-laboratory results presented in Ref.~\cite{ahern2024}. 
In addition to the stable 778\,nm optical output of previous lab-based demonstrations of the dual-color architecture~\cite{EmilyPaper}, this frequency standard incorporates an integrated optical frequency comb, allowing the generation of both stable microwave (10\,MHz, 1\,GHz) and optical (centered at 1550\,nm) outputs.  
We demonstrate a fractional frequency stability of $1.9\times10^{-13}$ at $\tau\,{=}\,1$\,s, reaching a landmark $3.5\times10^{-15}$ at $\tau\,{=}\,8000$\,s for any optical Rb frequency standard.% to the best of the authors' knowledge. 

%%%%%%%%%%%%%%%%%%%%%%%%%%%%%%%%%%%%%%%%%%%%%%%%%%%%%%%%%%%%%%%%%%%%%%%%%%%%%%%%%%%%%%%%%%%%%%%%%%%
%-------------------------------------------------------------------------------------------------%
%%%%%%%%%%%%%%%%%%%%%%%%%%%%%%%%%%%%%%%%%%%%%%%%%%%%%%%%%%%%%%%%%%%%%%%%%%%%%%%%%%%%%%%%%%%%%%%%%%%

\section{Experiment}
\label{sec:experiment}
The portable dual-color Rb two-photon frequency standard can be broadly split into three main interconnected systems, which will be described separately: the optical frequency standard's photonics; the optical frequency comb; and the supporting power, control, and oversight system.
The major components of these subsystems are shown schematically in Fig.\,\ref{fig:ClockInManyBoxes}b. 
%The frequency standard is housed in a 10 rack-unit (10U) (including fan and uninterruptible power supply (UPS); 7U without) portable 19-inch rack within several interconnected boxes as seen in Fig.\,\ref{fig:ClockInManyBoxes}c.
%The power consumption of the frequency standard without UPS is 340\,W, and weighs 45\,kg within its portable rack enclosure.
%The entire system can be lifted comfortably by two people, and is maneuverable by a single operator. 
The frequency standard is housed in mobile equipment rack as pictured in Fig.\,\ref{fig:ClockInManyBoxes}c. 
The rack has a 10 rack-unit (10U) capacity with an extended depth of 800\,mm.
The Rb frequency standard occupies 7U, with the remaining 3U used for a fan and uninterruptible power supply (UPS) unit. 
The power consumption of the frequency standard without UPS is 340\,W, and it weighs 45\,kg within its portable rack enclosure (without UPS unit).
The entire system can be lifted comfortably by two people with UPS unit installed, and is maneuverable by a single operator. 

%%%%%%%%%%%%%%%%%%%%%%%%%%%%%%%%%%%%%%%%%%%%%%%%%%%%%%%%%%%%%%%%%%%%%%%%%%%%%%%%%%%%%%%%%%%%%%%%%%%
%-------------------------------------------------------------------------------------------------%
%%%%%%%%%%%%%%%%%%%%%%%%%%%%%%%%%%%%%%%%%%%%%%%%%%%%%%%%%%%%%%%%%%%%%%%%%%%%%%%%%%%%%%%%%%%%%%%%%%%

\subsection{Two-Photon Frequency Standard Photonics}
\label{subsection:ExperimentOpticalSetup}
%The experimental setup of the frequency standard broadly follows the architecture presented in Refs.~\cite{ahern2024, EmilyPaper}, but is presented here briefly for completeness.
The Photonics subsystem is further subdivided into two sections: the laser preparation section, and the `physics package' containing the warm Rb vapor cell in which the two-photon transition occurs. 
Two telecommunications-band fiber lasers at 1552\,nm and 1560\,nm - once frequency-doubled to 776\,nm and 780\,nm respectively - drive the two-photon $5S_{1/2}\left(F=2\right)\rightarrow5D_{5/2}\left(F=4\right)$ transition as seen in Fig.\,\ref{fig:ClockInManyBoxes}a.
Successful excitation of the $5S_{1/2}\rightarrow5D_{5/2}$ transition yields 420\,nm fluorescence detected by the two photomultiplier tubes (PMTs) within the physics package.

The two lasers are combined using a wavelength division multiplexer (WDM), optically amplified with an erbium doped fiber amplifier (EDFA), then split back into separate frequencies with another WDM. 
Each laser is power modulated with an acousto-optic modulator (AOM) before frequency-doubling with second harmonic generators (SHGs) to generate the 776\,nm and 780\,nm photons to drive the $5S_{1/2}\rightarrow5D_{5/2}$ clock transition as seen in Fig.\,\ref{fig:ClockInManyBoxes}a.
The 776\,nm and 780\,nm light is combined with a 50:50 splitter before the two wavelengths are launched from the same collimator into the physics package to ensure co-linear propagation. 
Within the physics package, the dual-color light passes through a linear polarizer and optical wedge before being directed through the Rb cell. 
A mirror mounted at the rear of the Rb cell provides the counter-propagating optical fields required for quasi-Doppler-free excitation of the atoms. %retro-reflects the beam to facilitate counter-propagating excitation and a reduction in Doppler broadening while increasing the excitation rate of the two-photon transition.

The isotopically enhanced $^{87}$Rb vapor cell is heated by four embedded heater cartridges within the large thermal heat capacity stainless steel oven that holds the cell to an operating temperature of between $333\,\text{K}$ and $340\,\text{K}$ with a cell temperature stability of $\,1\mu\text{K}$.
Holding the cell at this temperature increases the vapor density of Rb atoms and thus fluorescence production.
The Rb cell temperature is maintained both passively via the large thermal mass of the stainless steel oven, with insulation provided by a 3D-printed Ultem (polytherimide) insulative layer and encapsulating $\mu$-metal shield, as well as an active temperature stabilization feedback loop.
The oven temperature is sensed using an embedded 4-wire measurement of a 10\,k$\Omega$ negative temperature coefficient (NTC) thermistor, which is used to feed back to the heater cartridges and maintain the nominal operating temperature. 
An additional out-of-loop four wire 10\,k$\Omega$ NTC thermistor is used to independently monitor oven temperature.

There are a number of frequency, power, and temperature control loops in the frequency standard.
The 420\,nm fluorescence from the 6$P_{3/2}$ decay path is used to frequency stabilize the 1552\,nm laser to the two-photon transition via frequency modulation (FM) spectroscopy at a frequency $f_\mathrm{mod}$ that is applied using the 1552\,nm AOM, with demodulation of the fluorescence signal detected on the physics package's PMTs producing a frequency discriminator for stabilization.
The power of each laser is monitored using an optical wedge to generate two reflected beams that are detected by photodiodes within the physics package: one for power control loops (Power PD in Fig.\,\ref{fig:ClockInManyBoxes}b) and the other for diagnostics.
As the two color components of the beam are spatially overlapped at this point, temporal separation of the two colors is used: the 780\,nm light is square wave amplitude modulated by the 1560\,nm AOM.
The square wave signal measured by the PD is filtered, demodulated, and fed into a proportional-integral (PI) feedback loop, which actuates upon the amplitude of the square wave modulation signal driving the 1560\,nm AOM, controlling the power of the 780\,nm light.
The 776\,nm power control is a DC stabilization loop actuating upon the 1552\,nm AOM that maintains the DC level of the Power PD at a defined set point.
As this is a DC measurement, this PD measurement includes the averaged optical power of the 780\,nm light, however minimal coupling has been observed between the two control loops due to their sufficiently different Fourier frequencies~\cite{EmilyPaper}.

The use of the 1552\,nm AOM to apply frequency modulation to the 1552\,nm laser for frequency stabilization creates unwanted laser power modulation at $f_\mathrm{mod}$, referred to as residual amplitude modulation (RAM). 
The phase and amplitude of this unwanted RAM contribution is detected on the same physics package PD (Power PD in Fig.\,\ref{fig:ClockInManyBoxes}b) as the two power control loops, with the signal isolated using an electronic filter.
This signal is used to produce a tone at the same frequency and amplitude but opposite phase, which is fed back into the the 1552\,nm AOM to suppress the RAM contribution. 
Excepting the physics package temperature control loop, all control loops are implemented with commercial-off-the-shelf field programmable gate array (FPGA) evaluation boards, overseen by the master control and automation computer described in Section\,\ref{subsection:PowerControlOversight}. 

%It must be noted that while the 1552\,nm and 1560\,nm laser frequencies are referenced to the two-photon clock transition, the individual laser frequencies are themselves unstable - it is only the sum of the laser frequencies that is stable.
While the sum of the frequencies of the two interrogating lasers is referenced to the clock transition, this does not constrain the individual laser frequencies.
To generate this stable optical output, a portion of the combined 1560\,nm and 1552\,nm light is split off after the EDFA prior to the second WDM, and is directed through a sum frequency generator (SFG), which generates the stable 778\,nm optical frequency standard output.
The integrated portable optical frequency comb with repetition rate stabilized to the frequency standard's stable 778\,nm optical output can also be used to generate stable microwave outputs at multiples of the repetition rate, or through further division, standard RF clock frequencies more compatible with electronic measurements~\cite{NicClockworks}.
The stable 778nm optical output can also compared to other optical Rb frequency standards that have an output at 778\,nm or by using an optical frequency comb to bridge a gap between two optical frequency standards of differing output wavelengths that fall within the spectral bandwidth of the comb, as demonstrated in Ref.~\cite{ahern2024}.

%%%%%%%%%%%%%%%%%%%%%%%%%%%%%%%%%%%%%%%%%%%%%%%%%%%%%%%%%%%%%%%%%%%%%%%%%%%%%%%%%%%%%%%%%%%%%%%%%%%
%-------------------------------------------------------------------------------------------------%
%%%%%%%%%%%%%%%%%%%%%%%%%%%%%%%%%%%%%%%%%%%%%%%%%%%%%%%%%%%%%%%%%%%%%%%%%%%%%%%%%%%%%%%%%%%%%%%%%%%
 
\subsection{Portable Optical Frequency Comb}
\label{subsection:ExperimentPortableComb}
A portable optical frequency comb is housed within the top 2U of the rack in Fig.\,\ref{fig:ClockInManyBoxes}c. 
The comb is based on the design developed at the National Institute of Standards and Technology (NIST)~\cite{Sinclair2015InvitedAA}.
The comb's carrier-envelope offset frequency ($f_\text{CEO}$) is stabilized with the traditional $f-2f$ self-referencing scheme~\cite{Sinclair2015InvitedAA}.
The repetition rate ($f_\text{rep}$) of the comb (200\,MHz) is stabilized to the 778\,nm stable frequency standard output via a phase-lock-loop measured using the Comb $f_\text{rep}$ PD in Fig.\,\ref{fig:ClockInManyBoxes}b.
The frequency of the 1560\,nm laser is stabilized to a mode of the optical frequency comb via another phase-lock-loop (using 1560\,nm Freq. PD in Fig.\,\ref{fig:ClockInManyBoxes}b), completing the stabilization of the frequency standard.
Selecting a different comb mode also allows modification of the intermediate state detuning ($\Delta_i$).

%%%%%%%%%%%%%%%%%%%%%%%%%%%%%%%%%%%%%%%%%%%%%%%%%%%%%%%%%%%%%%%%%%%%%%%%%%%%%%%%%%%%%%%%%%%%%%%%%%%
%-------------------------------------------------------------------------------------------------%
%%%%%%%%%%%%%%%%%%%%%%%%%%%%%%%%%%%%%%%%%%%%%%%%%%%%%%%%%%%%%%%%%%%%%%%%%%%%%%%%%%%%%%%%%%%%%%%%%%%

\subsection{Automation and Oversight}
\label{subsection:PowerControlOversight}
The dual-color, two-photon Rb optical frequency standard, including portable optical frequency comb, is capable of fully autonomous cold-start operation.
It performs warm-up and stabilization of all feedback loops described previously as controlled by a custom Python master control program running on the oversight computer housed in the control and oversight rack unit (under the blue Photonics box in Fig.\,\ref{fig:ClockInManyBoxes}c).
In addition, the frequency standard has some error handling capability, as well as fail-safe shutdown procedures.
Remote operation and monitoring is possible by connecting to the secure network-enabled oversight computer. 
The unit is powered by 240VAC with an uninterruptible power supply (UPS) providing electrical filtering and approximately one hour of battery-only operation. 
The frequency standard has seen successful remote operation in movement between buildings, in the back of a moving vehicle, and aboard the deck of a naval platform during exercises on the Pacific Ocean after international transport~\cite{ahern2024}. 

%%%%%%%%%%%%%%%%%%%%%%%%%%%%%%%%%%%%%%%%%%%%%%%%%%%%%%%%%%%%%%%%%%%%%%%%%%%%%%%%%%%%%%%%%%%%%%%%%%%
%-------------------------------------------------------------------------------------------------%
%%%%%%%%%%%%%%%%%%%%%%%%%%%%%%%%%%%%%%%%%%%%%%%%%%%%%%%%%%%%%%%%%%%%%%%%%%%%%%%%%%%%%%%%%%%%%%%%%%%

\begin{figure}[t]
    \centering
    \includegraphics[width=\columnwidth]{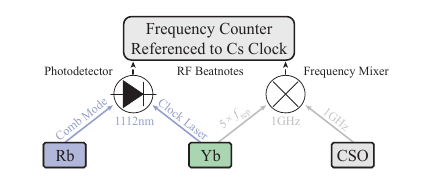} 
    \caption{Schematic of the pre-trial comparison system used to measure the frequency stability of the frequency standard, with microwave outputs (light grey), optical outputs (blue) and down-mixed beat notes (dashed black lines). Post trial, the microwave output from the Yb standard's comb was replaced by the same output from the Rb standard's comb. The optical comparison was not measured post trial.}
    \label{fig:MeasurementSummary}
\end{figure}

\subsection{Stability Measurements}
\label{subsection:Measurement}
In order to characterize the stability of the frequency standard, two measurement methods were employed as shown in Fig.\,\ref{fig:MeasurementSummary}. 
Prior to RIMPAC-2022, two simultaneous measurements of three frequency references were performed. 
Firstly, an optical beat note was generated between the portable Rb frequency standard and a second portable optical frequency standard based upon interrogation of a thermal ytterbium (Yb) vapor ~\cite{ahern2024}.
This was performed at double the wavelength of the Yb transition ({1112\,nm), which is directly available from the Yb standard, and the nearest accessible comb mode from the Rb standard's optical frequency comb~\cite{ahern2024}. 
Secondly, a beat note was measured between the fifth harmonic of the Yb frequency standards' optical frequency comb repetition rate, $5\times f_\text{rep}\approx1$\,GHz, and the 1\,GHz output from a lab-based cryogenic sapphire oscillator (CSO)~\cite{Tobar2006}.
Both beat notes were recorded on a commercial zero dead-time frequency counter referenced to a Microchip 5071A microwave cesium beam clock~\cite{ Kramer:01, Kramer:04}.  
These two simultaneous measurements can be used to extract the performance of the portable Rb standard from the ensemble.
Upon its return from the ship-borne exercise another measurement was made.
For this measurement, the fifth harmonic of the portable Rb standard's optical frequency comb was compared to the same CSO as used previously.
%A different non-zero dead time frequency counter was used for the post-test measurement, and without the second optical frequency standard. 
%It was noted upon return from RIMPAC-2022 that the performance of the portable Rb standard over all time scales had degraded slightly due to a known mechanical issue, which is discussed in the next section.
%However, the frequency stability measurement setup utilized upon return was the same measurement setup as that used by \cite{EmilyPaper}, which demonstrates that the performance degradation of the portable frequency standard upon return was not due to a change in the specifics of the stability measurement.

%%%%%%%%%%%%%%%%%%%%%%%%%%%%%%%%%%%%%%%%%%%%%%%%%%%%%%%%%%%%%%%%%%%%%%%%%%%%%%%%%%%%%%%%%%%%%%%%%%%
%-------------------------------------------------------------------------------------------------%
%%%%%%%%%%%%%%%%%%%%%%%%%%%%%%%%%%%%%%%%%%%%%%%%%%%%%%%%%%%%%%%%%%%%%%%%%%%%%%%%%%%%%%%%%%%%%%%%%%%

%\section{Results}
%\label{sec:Results}
\section{Results \& Discussion}
\label{sec:Results}

\begin{figure}[htbp]
    \centering
    \includegraphics[width=\columnwidth]{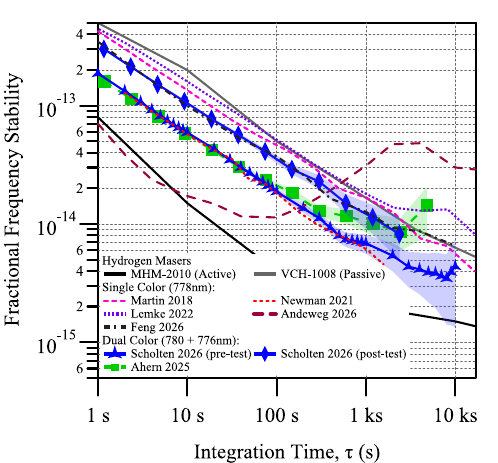} 
    \caption{Comparison of two-photon frequency standard stabilities (Allan deviations~\cite{Rutman1991}). 
    All presented data are without any drift removal. Five single-color frequency standards are presented~\cite{Martin2018,Newman2021, Lemke, Andeweg2026, Feng2026} (pink, red, purple, maroon, and black dashed lines respectively), along with a previously presented stability of this dual-color approach~\cite{EmilyPaper} (closed green squares). 
    The reported performance of two commercial hydrogen masers are presented for comparison: the Microchip MHM-2010 active hydrogen maser (black solid line), and Veremya-CH VCH-1008 19" rack mounted passive hydrogen maser. 
    The performance of the portable system prior to overseas transport (closed blue caltrops, pre-test) is on par with the single-color designs and surpasses performance of all previously reported demonstrations between 2000\,s and 10,000\,s integration time. 
    %This represents a significant improvement on previous results for the dual-color system at medium to long integration times. Post transport (post-test) the performance is reduced, however still comparable with previous demonstrations (closed blue diamonds).
    %We believe the performance degradation is due to a mechanical design flaw in the physics package as described in Ref.~\cite{EmilyPaper}, which shares the same physics package design.
    }
    \label{fig:ClockComparison}
\end{figure}

The measured performance of the frequency standard in the laboratory is presented in Fig.\,\ref{fig:ClockComparison}. 
Prior to international transport, we show a fractional frequency stability of $1.9{\times}\,10^{-13}$ at $\tau\,{=}\,1$\,s, averaging down with shot-noise limited performance as $1/\sqrt\tau$ to $3.5\times10^{-15}$ at $\tau\,{=}\,8000$\,s.
After the international field trial, the stability was measured to be $3.0{\times}\,10^{-13}$ at $\tau\,{=}\,1$\,s, averaging down to $8.2\,\times\,10^{-15}$ at $\tau\,{=}\,2400$\,s.
The pre-transport performance (closed blue caltrops, pre-test in Fig.\,\ref{fig:ClockComparison}) represents a significant improvement over previously presented laboratory-based dual-color two-photon frequency standard stabilities at $\tau>200$\,s~\cite{Perrella2019, EmilyPaper}.
Furthermore, the results are comparable to lab-based single-color, two-photon frequency standard configurations also presented in Fig.\,\ref{fig:ClockComparison}~\cite{Martin2018, Newman2021, Lemke} until $\tau=2000$\,s, after which literature values on comparable timescales for the single color architecture do not exist or have worse performance.
%The performance of the portable system prior to overseas transport is on par with the single-color architecture demonstrations.
%Additionally, this matches the performance of the previously published results for the dual-color lab-based system under similar operating conditions, and is a significant improvement at integration times of longer than 1000\,s~\cite{EmilyPaper, Perrella2019, ahern2024}. 

Post transport and operation at RIMPAC-2022 (post-test in Fig.\,\ref{fig:ClockComparison}) the performance of the portable Rb frequency standard is reduced over all time scales, however it is still comparable with previous demonstrations of lab-based single-color standards~\cite{Martin2018, Lemke} without the need for helium permeation mitigation pre-treatment of the rubidium cell~\cite{Feng2026}.
We believe the performance degradation post-trial is due to the physics package mechanical design flaw affecting beam alignment as described in Ref.~\cite{ahern2024}, which shares the same physics package design.
For future demonstrations, the mechanical alignment instability of the interrogating beams in the current physics package design should be addressed to prevent the observed performance degradation after transport of the system. 
The single-color lab demonstration presented in Ref.~\cite{Andeweg2026} surpasses the performance of the standard presented here between $\tau = 1$\,s to $\tau = 200$\,s when the active AC Stark shift compensation is operational, as this allows for higher optical powers to be used with a corresponding improvement in short-term performance~\cite{Margolis2010}.  

Both pre- and post-test performance measurements in Fig.\,\ref{fig:ClockComparison} surpass the typical performance of a comparable 19" rack-mounted commercial passive hydrogen maser (Veremya-CH VCH-1008) over all measured timescales, demonstrating the potential capabilities of portable optical atomic clocks for drop-in replacement of such systems with future size, weight, and power (SWaP) reductions.
Future work is split between addressing identified limits to long term stability (namely pressure shift, 780\,nm light shift, and the effects of residual amplitude modulation) through improved control systems, and SWaP reduction through bespoke engineering solutions such as the incorporation of photonic integrated circuits~\cite{EmilyPaper, Maurice2020, Knappe05, Martinez2023, LOSEV2015242, MEMSStrathclyde, Tran2022, LongCell}.
%A full characterization of noise sources and their effects on the stability of a lab-based variant of the portable standard described here is presented in Ref.~\cite{EmilyPaper}. 

%\section{Discussion}
%\label{sec:Discussion}

%IMPORTANT NOVELTY POINTS ARE SWAP AND PORTABILITY, NEED TO PUT IN CONTEXT OF OTHER PORTABLE CLOCK DEMONSTRATIONS. WHAT DOES A PORTABLE CLOCK ENABLE: What does a portable clock enable? Fundamental Physics tests (Katori) or Defence Applications (Our Nature Comms’s & the Vector Atomic paper) DISTINGUISH FROM EMILY AND THE NAT COMMS PAPER. 

The performance of the frequency standard presented here confirms that portable optical atomic clocks, with integrated optical frequency combs, are rapidly approaching suitability for real-world use cases.
With additional development and SWaP reduction, such devices are on the cusp of becoming commercial off the shelf components suited for drop-in replacements of microwave systems whilst providing improved performance for applications such as providing reliable positioning, navigation, and timing for vessels and vehicles, commercial synchronization and scientific synchronization of distributed sensors, measurements of relativistic geodesy, and tests of fundamental physics amongst other applications.

\section{Conclusion} \label{sec:conclusion}
We demonstrate the first fully autonomous and portable optical frequency standard based on a dual-color excitation of the $5S_{1/2}\rightarrow5D_{5/2}$ two-photon transition in $^{87}$Rb.
The device was fielded at the RIMPAC 2022 international naval exercise~\cite{ahern2024}, starting and operating autonomously for the duration of the exercise, and performing as a similar level upon its return.
Additionally, the standard demonstrates comparable frequency stability in both the optical and radio frequency domains with an integrated optical frequency comb, as demonstrated by optical and microwave comparisons.
The standard exhibits a fractional frequency stability of $1.9\times10^{-13}$ at an integration time of 1\,s, reaching $3.5\times10^{-15}$ at 8000\,s, constituting the best performance of an optical Rb frequency standard of either the single or dual-color designs to date. 
This portable frequency standard represents an important step towards a low SWaP dual-color, two-photon Rb frequency standard, and optical frequency standards operating autonomously outside the laboratory.
%Furthermore, these results are achieved with the first fully portable Rb system with outputs in both the optical and radio frequency domains.

%%%%%%%%%%%%%%%%%%%%%%%%%%%%%%%%%%%%%%%%%%%%%%%%%%%%%%%%%%%%%%%%%%%%%%%%%%%%%%%%%%%%%%%%%%%%%%%%%%%
%-------------------------------------------------------------------------------------------------%
%%%%%%%%%%%%%%%%%%%%%%%%%%%%%%%%%%%%%%%%%%%%%%%%%%%%%%%%%%%%%%%%%%%%%%%%%%%%%%%%%%%%%%%%%%%%%%%%%%%

\acknowledgments
%This research is supported by the Commonwealth of Australia Defence Science and Technology Group (DST Group). 
%We would particularly like to thank DST Group staff including Scott Foster, Joanne Harrison, Ben Sparkes, David Bird, and Anthony Szabo for their support of the Adelaide University (formerly University of Adelaide) portable atomic clocks team and the Rb two-photon frequency standard's development.
This research is supported by the by the Australian Government and Commonwealth of Australia Defence Science and Technology Group (DST Group) through the Next Generation Technologies Fund.
We would particularly like to thank DST Group staff including Scott Foster, Joanne Harrison, Ben Sparkes, David Bird, and Anthony Szabo for their support of the University of Adelaide portable atomic clock team and the rubidium clock's development.
We acknowledge support from the U.S. AFOSR AOARD FA2386-19-1-4054 and FA2386-20-1-4032.
The authors thank the Optofab node of the Australian National Fabrication Facility (ANFF), which utilizes Commonwealth and South Australian State Government funding. 
In particular, Evan Johnson, Lijesh Thomas, Alastair Dowler, and Alson Ng.
This research was conducted by the Australian Research Council Centre of Excellence in Optical Microcombs for Breakthrough Science (project number CE230100006) and funded by the Australian Government.
S. K. S is the recipient of an Australian Research Council Australian Early Career Researcher Industry Fellowship (project number IE240100056) funded by the Australian Government.
This project and publication is supported in part by the Australian Government Department of Education through Australia’s Economic Accelerator Program.
The authors thank members of NIST for their guidance in constructing the optical frequency comb, particularly Tara Fortier, Laura Sinclair, Esther Baumann, and Ian Coddington. 
We acknowledge Benjamin White for their contributions to the lab demonstration of the Rb two-photon frequency standard, on which the architecture from this work was based.
The work presented is covered by patent US 10353270 B2.
%%%%%%%%%%%%%%%%%%%%%%%%%%%%%%%%%%%%%%%%%%%%%%%%%%%%%%%%%%%%%%%%%%%%%%%%%%%%%%%%%%%%%%%%%%%%%%%%%%%
%-------------------------------------------------------------------------------------------------%
%%%%%%%%%%%%%%%%%%%%%%%%%%%%%%%%%%%%%%%%%%%%%%%%%%%%%%%%%%%%%%%%%%%%%%%%%%%%%%%%%%%%%%%%%%%%%%%%%%%

\bibliography{mainBib}

\end{document}